\documentclass{ceurart}
\usepackage{booktabs}   
\usepackage{tabularx}   
\usepackage{amsmath}    
\usepackage{graphicx}   
\usepackage{placeins}   
\usepackage{threeparttable}   
\usepackage{hyperref}

\begin{document}

\copyrightyear{2025}
\copyrightclause{Copyright for this paper by its authors.
  Use permitted under Creative Commons License Attribution 4.0
  International (CC BY 4.0).}

\conference{AIC 2025: The 10th International Workshop on Artificial Intelligence and Cognition (held as part of ECAI 2025). October 25-26, 2025. Bologna, Italy}

\title{Meaning in Order, Order in Meaning: \\ Semantic R-precision for Keyphrase Evaluation}

\author[1, 2]{Shamira Venturini}[%
orcid=0009-0000-8202-1320, 
email=shamira.venturini@h-ka.de, 
]
\cormark[1]
\address[1]{ILIN - Institute for Learning and Innovation in Networks, Karlsruhe University of Applied Sciences, Moltkestraße 30, 76133 Karlsruhe, Germany} 

\author[1]{Steffen Kinkel}[%
orcid=0000-0001-6154-136X, 
]
\address[2]{Karlsruhe Institute of Technology, Kaiserstraße 12, 76131 Karlsruhe, Germany} 

\cortext[1]{Corresponding author.}

\begin{abstract}
Evaluating the quality of automatically generated keyphrases remains a complex challenge. Traditional metrics either rely on exact lexical matching or consider semantic similarity while ignoring prediction ranking, both of which misalign with how humans judge informativeness and relevance. We introduce Semantic R-Precision (SemR-p), a novel evaluation metric that integrates semantic similarity into the rank-aware R-Precision framework. Designed from a human-centric perspective and inspired by Information Retrieval metrics, SemR-p rewards semantically relevant keyphrases that appear early in the output list. We conducted extensive analyses to assess its semantic sensitivity, ranking awareness, and discriminative power across models and datasets. The results suggest that SemR-p offers a complementary lens for evaluating keyphrase predictions, helping to better reflect user-centred notions of relevance alongside traditional lexical and semantic matching metrics.
\end{abstract}


 \maketitle

\section{Introduction}
\label{sec:introduction}

Automatically generated keywords are increasingly used to summarise and retrieve documents. Evaluating the quality of these generated keyphrases is typically automated, but current evaluation methods often fall short of capturing what humans actually perceive as meaningful, relevant or necessary \cite{Cai_Leckner_Björklund_2025}. The standard evaluation process involves comparing a model’s predicted keyphrases against a set of human-annotated reference keyphrases for the same document. Metrics then score the prediction list based on how well it overlaps with the reference list. A major limitation of this approach is its reliance on exact lexical matches, which overlooks semantically equivalent but differently worded keyphrases.

Recent approaches, like BERTScore \cite{zhang-etal-2019-BERTscore}, address this limitation by measuring semantic similarity between predictions and references using contextual embeddings from large language models. However, they do not evaluate the order in which they are presented, even though ranking can strongly influence perceived informativeness and user expectations, as users tend to focus more on top-ranked results due to well-known cognitive biases such as the primacy effect \cite{Cai_Leckner_Björklund_2025}. 

To address these limitations, we propose \textbf{Semantic R-Precision} (SemR-p) \footnote{Code is available as a \href{https://github.com/shamira-venturini/semantic-r-precision}{standalone implementation} and integrated into our \href{https://github.com/shamira-venturini/KPEval-SemR-p}{fork of the KPEval toolkit} from the Natural Language Processing @UCLA repository, released under the MIT License.}, a novel evaluation metric that integrates semantic similarity scoring into the rank-sensitive R-Precision framework \cite{zesch-gurevych-2009-approximate, manning2008introduction}, a novel metric integrating semantic similarity scoring within the rank-sensitive R-Precision framework. Grounded in human cognitive principles and inspired by evaluation practices in Information Retrieval, our approach shifts from exact lexical matching to assessing semantic equivalence, while preserving a user-centred perspective. This enables a more cognitively plausible assessment of keyphrase predictions, reflecting how users actually judge informativeness and relevance. While our long-term goal is to develop evaluation methods that closely align with human judgment, this work focuses on establishing SemR-p’s computational formulation and empirical behaviour.

This paper introduces SemR-p and presents a preliminary comprehensive evaluation to demonstrate its properties, utility and computational inner workings. Our contributions include (1) a formal definition (Section \ref{sec:semrp_definition}); (2) an extensive empirical analysis across models and datasets, and (3) qualitative case studies. We begin by reviewing related work (Section \ref{sec:related_work}) before detailing our methods (Section \ref{sec:methodology}) and presenting our results (Section \ref{sec:results}). We conclude with a discussion of findings, limitations and future work (Sections \ref{sec:discussion} and \ref{sec:limitations}).


\section{Related Work}
\label{sec:related_work}

The evaluation of automatically generated keyphrases is a challenging task, with different metrics emphasising distinct aspects of quality (exact correspondence, semantic similarity, saliency, etc). The \textit{Cognitive Model of Information Retrieval }proposed by Sutcliffe and Ennis \cite{Sutcliffe1998} characterises human information evaluation as a subjective process that is sensitive to both meaning and presentation order. However, traditional automated evaluation methods often fail to account for these human-centric factors \cite{Firoozeh_Nazarenko_Alizon_Daille_2020}. In this section, we review major evaluation paradigms through this lens, highlighting a disconnect between how metrics are designed and how humans perceive keyphrase quality.

\subsection{Exact lexical matching metrics}
The most common evaluation paradigm relies on exact lexical matching, typically using Precision, Recall, and the F1-score to measure the overlap between the set of predicted keyphrases $\hat{Y}$ and the reference set $Y$. Consider a document where the references are $Y =$ \{\texttt{neural networks}, \texttt{machine translation}, \texttt{attention mechanism}\}. A model might predict the list $\hat{Y} =$ \{\texttt{machine translation}, \texttt{deep learning}, \texttt{attention mechanism}\}. Classic $F1$ would only find two matches (\texttt{machine translation} and \texttt{attention mechanism}), ignoring the strong semantic relationship between \texttt{neural networks} and \texttt{deep learning}. Human cognition focuses on concepts that can be expressed through various lexical forms (e.g., synonyms, paraphrases), and studies of human-assigned keyphrases have shown that different annotators often use such variations to describe the same core ideas \cite{kim-etal-2010-evaluating} and might use words not even appearing in the text \cite{meng-etal-2017-deep}. Despite being counter-intuitive, a comprehensive survey revealed that two-thirds of studies from 2017–2023 still adopted these metrics \cite{wu-etal-2024-kpeval}.

\subsection{Approximate matching metrics}
To make automatic keyphrase evaluation less strict, Zesch and Gurevych~\cite{zesch-gurevych-2009-approximate} adapted the R-Precision metric from information retrieval using an approximate matching strategy. The original R-Precision framework computes the proportion of relevant documents found within the top $R$ retrieved results, where $R$ is the total number of relevant documents for the query \cite{manning2008introduction}. They modified this metric by granting full credit for close variants, such as morphological forms (\textit{MORPH}: (P)~ \texttt{machines} and (R)~ \texttt{machine}), for predicted phrases that included a reference phrase (\textit{INCLUDES}: (P)~\texttt{advanced machine learning} and (R)~\texttt{machine learning}), or when a prediction was contained within a reference (\textit{PARTOF}: (P) \texttt{information retrieval} – (R) \texttt{information}). However, when they asked human annotators to judge how meaningful these partial matches were, they found low agreement, especially for cases where one phrase simply included the other as a substring.

Kim et al.~\cite{kim-etal-2010-evaluating} proposed a modification to R-Precision that weights matches based on the position of words within a keyphrase. In English noun phrases, the head noun typically appears at the rightmost position and carries the main semantic and syntactic weight of the phrase. To capture this, their modified R-Precision assigns weights to each word based on its position: the rightmost word, the head noun, is assigned the highest weight of 1, while weights decrease for words further to the left. 
Partial matches that correctly overlap with words closer to the head noun receive greater credit. 
Kim et al. \cite{kim-etal-2010-evaluating} showed that this weighted scoring scheme correlates better with human judgments on partially overlapping keyphrases compared to unweighted metrics that treat all matching words equally. However, while this method improves evaluation by considering syntactic structure, it still relies on word overlap, and does not capture deeper semantic relationships, which can be better modelled by modern embedding-based approaches.

\subsection{Semantic Similarity Metrics}

To move beyond lexical similarity altogether, researchers have proposed metrics that focus on semantic similarity. In order to capture meaning, these metrics typically use contextual embeddings, which are vector representations of text from pre-trained language models. A similarity score is then computed by comparing the embeddings of predicted and reference phrases. The most prominent example is BERTScore \cite{zhang-etal-2019-BERTscore}, which compares token-level vectors via cosine similarity to calculate aggregate Precision, Recall, and F1 scores. Because BERTScore operates at the token level, it may fail to capture the holistic meaning of a keyphrase. For example, while \texttt{machine} and \texttt{translation} are common words, \texttt{machine translation} as a phrase refers to a specific concept. Evaluating tokens separately may miss that nuance.

To address this, Wu et al. \cite{wu-etal-2024-kpeval} proposed semantic variants of Precision, Recall and F1 (SemP, SemR, SemF1) which compare mean-pooled embeddings of entire phrases. These methods are better aligned with the conceptual unit of a keyphrase, offering a more semantically coherent basis for comparison. For example, it would better capture the semantic similarity between \texttt{artificial neural network} and \texttt{deep learning model}, since both describe similar ideas despite differing in surface form.
However, as traditional Precision, Recall and F1, they treat predictions as unordered sets, disregarding the order of presentation. Recent user-centred studies show that the perceived quality of keyphrases depends not only on their content but also on their presentation order, aligning with known attentional biases toward earlier list items \cite{Cai_Leckner_Björklund_2025}.

\subsection{Rank-aware Metrics}

Evaluation paradigms from Information Retrieval, by contrast, place strong emphasis on this ranking quality. Generally, they assign greater weight to relevant items found at higher positions (Average Precision (AP) \cite{manning2008introduction}) in a results list, often by applying a positional discount to the value of items as their rank decreases (Normalised Discounted Cumulative Gain (NDCG) \cite{Jrvelin2002} and Mean Reciprocal Rank (MRR) \cite{manning2008introduction}). For example, if \texttt{machine translation} appears at rank 1 and is a relevant keyphrase, it contributes more to the final score than if it appears at rank 5. Metrics like NDCG apply a logarithmic discount to account for this drop in user attention with lower-ranked items. However, when applied to keyphrase evaluation, the determination of an item's relevance is still typically based on exact or approximate lexical matching, thus failing to capture semantic equivalence.

One additional challenge is the scope of the evaluation: fixed cut-offs (e.g., top 5) ignore that the number of relevant keyphrases varies across documents. However, evaluating full lists can reward over-generation \cite{yuan-etal-2020-one}. R-Precision offers a compromise by scoring only the top $R$ predictions (where $R = |Y|$), but it still relies on surface-form matching. 

This review highlights a clear gap: the need for evaluation metrics that integrate robust semantic understanding within a rank-sensitive, human-aligned framework. Future methods must capture both the conceptual meaning of keyphrases and their cognitive impact on users, in order to support more trustworthy and interpretable evaluation of keyphrase generation systems. 

\section{Semantic R-precision}
\label{sec:semrp_definition}

We build on the R-Precision framework \cite{zesch-gurevych-2009-approximate}, which evaluates only the top-R predictions ($R = |Y|$), thereby modelling user attention \cite{Cai_Leckner_Björklund_2025, Jansen2006} and adapting the evaluation scope to document complexity. Our core novelty is the integration of a robust semantic similarity assessment within this rank-sensitive structure. 
Since humans process keyphrases as holistic concepts rather than as bags of tokens, we employ phrase-level embeddings. The relevance score for each prediction is determined by a two-part logic that balances precision with semantic flexibility. First,  an exact stem match is rewarded with a full score. For non-matches, we calculate a semantic score by averaging the similarity to the top-$k$ most similar references. This helps the metric choose the best conceptual fit and reduces noise from unrelated data, similar to how humans judge relevance by focusing on salient concepts \cite{rosch1975cognitive, ingwersen2005turn, Cai_Leckner_Björklund_2025}.

SemR-p evaluates the top $R$ predictions from the ordered list $\hat{Y} = (\hat{y}_1, \dots, \hat{y}_m)$, where $R = |Y|$ is the number of reference keyphrases in the set $Y = \{y_1, \dots, y_r\}$. For each of the top $R$ predicted keyphrases $\hat{y}_i$:
\begin{enumerate}
    \item \textbf{Exact Stem Match Check:} If the stem of $\hat{y}_i$ exactly matches the stem of any reference keyphrase $y_j \in Y$, the score for $\hat{y}_i$ is $score(\hat{y}_i) = 1.0$. where $stem(\cdot)$ is a stemming function.
    
    \item \textbf{Semantic Similarity Score:} If $\hat{y}_i$ does not have an exact stem match in $Y$, its score is calculated based on its semantic similarity to the reference keyphrases. Specifically, the cosine similarity between phrase embeddings $emb(\hat{y}_i)$ and the embeddings of all $y_j \in Y$ is computed.  The score is then the average similarity to the top-$k$ most similar reference keyphrases:
    \begin{equation}
        score(\hat{y}_i) = \frac{1}{k} \sum_{j=1}^{k} \cos\_sim(emb(\hat{y}_i), emb(y'_j))
    \end{equation}
    where $y'_j$ are the $k$ reference keyphrases in $Y$ most similar to $\hat{y}_i$ based on cosine similarity of their embeddings, and $k$ is a tunable parameter.
    \item \textbf{Final Metric Calculation:} The overall SemR-p score for the document is the average of the individual scores obtained for each of the top $R$ predictions:
    \begin{equation}
        SemR-p = \frac{1}{R} \sum_{i=1}^{R} score(\hat{y}_i)
        \label{eq:semrp}
    \end{equation}
\end{enumerate}

This formulation allows SemR-p to flexibly reward both exact matches and semantically similar predictions within the top $R$ ranks, maintaining a rank-aware evaluation framework aligned with practical use cases where top results are prioritised. To illustrate this calculation, Table~\ref{tab:worked_example} provides a step-by-step example.

Suppose the reference keyphrases for a document are: $Y$ = \{\texttt{neural networks},\ \texttt{machine translation},\ \texttt{deep learning}\}. And the model's predictions are:
$\hat{Y}$ = (\texttt{machine translation},\ \texttt{AI systems},\ \texttt{neural computation},\ \texttt{language modelling},\ \texttt{translation quality}). Since $R$ = $Y$ = 3, we evaluate only the top 3 predictions.

\begin{table*}[htbp!]
\centering
\caption{Step-by-step calculation of \textit{SemR-p} for a worked example.}
\label{tab:worked_example}
\begin{tabular}{l l p{4.5cm} p{3cm} r}
\toprule
\textbf{Top-R Prediction} & \textbf{Match Type} & \textbf{Top-k Similarities} & \textbf{Calculation} & \textbf{Score} \\
\midrule
\textbf{machine translation} & Exact & \textit{-} & $-$ & \textbf{1.00} \\
\addlinespace
\textbf{AI systems} & Semantic & 
    \begin{tabular}{@{} p{3.5cm} r @{}} 
        \texttt{deep learning} & 0.72 \\
        \texttt{neural networks} & 0.68 \\
        \texttt{machine translation} & 0.61 \\
    \end{tabular} & 
    \footnotesize{$(0.72+0.68+0.61)/3$} & \textbf{0.67} \\
\addlinespace
\textbf{neural computation} & Semantic & 
    \begin{tabular}{@{} p{3.5cm} r @{}}
        \texttt{neural networks} & 0.76 \\
        \texttt{deep learning} & 0.66 \\
        \texttt{machine translation} & 0.60 \\
    \end{tabular} & 
    \footnotesize{$(0.76+0.66+0.60)/3$} & \textbf{0.67} \\
\bottomrule
\end{tabular}
 \begin{tablenotes}[para,flushleft] 
            \[
            \textit{SemR-p} = \frac{1.00 + 0.67 + 0.67}{3} \approx \textbf{0.78}
            \]
\end{tablenotes}
\end{table*}
\FloatBarrier %
\section{Methodology}
\label{sec:methodology}
To evaluate the empirical behaviour and discriminative power of SemR-p, we designed a set of complementary analyses, each targeting a specific property desirable in a keyphrase evaluation metric. Our overarching goal was to assess whether SemR-p is (i) sensitive to semantic similarity, (ii) aware of prediction ranking, and (iii) able to distinguish between systems and datasets in meaningful ways.

This section is structured as follows: we first describe the experimental setup, including datasets, model predictions, baseline metrics, and implementation details. We then outline five analyses designed to evaluate SemR-p along the above dimensions: an ablation study on its semantic scoring parameter, a statistical sensitivity analysis, a system-level ranking agreement test, a latent factor analysis, and a qualitative case study evaluation.
\subsection{Experimental Setup}
\label{sec:experimental_setup}

Our experiments build on the KPEval toolkit \cite{wu-etal-2024-kpeval}, which provides reference annotations and pre-computed model predictions for two benchmark datasets: \textit{kp20k} \cite{meng-etal-2017-deep} and \textit{kptimes} \cite{gallina-etal-2019-kptimes}.

\paragraph{Data Sources and Models} The two benchmark datasets are \textit{kp20k} \cite{meng-etal-2017-deep} (scientific abstracts in computer science) and \textit{kptimes} \cite{gallina-etal-2019-kptimes} (news articles). From the 21 models available in KPEval, we curated a representative subset of eight, summarised in Table~\ref{tab:models}, spanning unsupervised and supervised extraction, supervised generation, and large language model-based generation. This selection ensures a broad coverage of methodological paradigms. Predictions from each model were evaluated on 1,000 documents per dataset.

\begin{table}[h] 
\caption{Selected keyphrase extraction and generation models and their methodologies.}
\label{tab:models} 
\centering
\begin{tabular}{ll} 
\toprule
\textbf{Model} & \textbf{Methodology} \\
\midrule
\multicolumn{2}{c}{\textit{Unsupervised Extraction}} \\ 
\texttt{tf-idf} \cite{jones-1972-tf-idf} & Statistical \\
\texttt{textrank} \cite{mihalcea-tarau-2004-textrank} & Graph-based Ranking \\
\texttt{multipartiterank} \cite{boudin-2018-unsupervised} & Graph-based Ranking \\
\midrule
\multicolumn{2}{c}{\textit{Supervised Extraction}} \\ 
\texttt{bert+crf} \cite{wu-etal-2022-bert+crf} & Sequence Labeling \\
\texttt{hypermatch} \cite{song-etal-2022-hyperbolic} & Hyperbolic Relevance Matching \\
\midrule
\multicolumn{2}{c}{\textit{Supervised Generation}} \\ 
\texttt{settrans} \cite{ye-etal-2021-one2set} & Set Prediction \\
\texttt{sciBART-large-OAGKX} \cite{wu-etal-2022-bert+crf} & Sequence-to-Sequence \\
\midrule
\multicolumn{2}{c}{\textit{LLM-based Generation}} \\ 
\texttt{GPT-3.5} & Five-Shot Prompting \\
\bottomrule
\end{tabular}
\end{table} 

\paragraph{Baseline Metrics} To contextualise the behaviour of SemR-p, we compared it against ten established metrics from prior work (Section~\ref{sec:related_work}): lexical exact-match (F1@M); rank-aware (NDCG, AP); approximate match (R-Precision); and semantic metrics (SemP, SemR, SemF1, and BERTScore variants: -P, -R, -F1). These metrics span distinct evaluation dimensions and serve as baselines for all subsequent analyses.

\paragraph{Implementation Details} For the semantic component of SemR-p and other baselines requiring phrase embeddings (with the exception of BERTscore), we used the \texttt{uclanlp/keyphrase-mpnet-v1} Sentence Transformer model \cite{reimers-gurevych-2019-sentence}. Phrase representations were obtained via mean-pooling of the model's output token embeddings. This is a standard and effective technique for deriving a single vector that captures the holistic meaning of a phrase. It balances representational quality with computational efficiency. We selected this specific model as it was fine-tuned on 1.04 million keyphrases and was empirically shown by Wu et al. \cite{wu-etal-2024-kpeval} to correlate well with human judgments of keyphrase similarity, making it a strong and validated choice for our experiments. 

For the lexical component, the `stem()` function in our formulation was implemented using the Porter Stemmer \cite{Porter2006}, consistent with Wu et al.'s implementation of SemP, SemR, SemF1 \cite{wu-etal-2024-kpeval}. Stemming allows for rewarding core lexical concepts without penalising minor morphological variations while also acting as an efficient computational shortcut.

All scores were computed per document and using $k \in {1, 2, 3}$ to explore semantic neighbourhood size effects, forming the basis for subsequent analyses.

\subsection{Evaluation Protocol}
\label{sec:eval_protocol}

To comprehensively evaluate the behaviour and utility of SemR-p, we conducted five analyses targeting its semantic validity, ranking sensitivity, robustness to parameter tuning, and interpretability. These analyses are detailed in the following subsections.

\paragraph{Ablation Study} 

The parameter $k$ in the SemR-p formulation determines how many of the most similar references are considered when calculating the semantic score for non-exact matches. Understanding its effect is crucial for interpreting the metric and establishing a robust default setting. We focused on a small value range, $k \in \{1, 2, 3\}$, as this principled choice focuses the evaluation on the most relevant references, reducing potential noise from semantically distant matches and enhancing interpretability.

\textit{Research Question:} How does varying the parameter $k$ affect SemR-p’s scoring behaviour and its alignment with other metrics across different datasets? 

\textit{Method:} We varied $k \in {1, 2, 3}$ and computed SemR-p scores across all documents. The analysis comprises three parts:

\begin{enumerate}
    \item Distribution of absolute SemR-p scores as $k$ changes.
    \item Spearman correlation of SemR-p (at each $k$) with semantic metrics (SemP/R/F1, BERTScore-P/R/F1).
    \item Correlation with lexical and rank-based metrics (F1@M, NDCG, AP, R-Precision) at the model level.
\end{enumerate}

We used paired statistical tests: Wilcoxon signed-rank test and Cliff's Delta effect sizes to assess the significance and magnitude of any observed changes. 

\paragraph{Sensitivity to Model Differences} 
To validate whether SemR-p can distinguish between systems and domains, we tested its statistical sensitivity to different models and datasets.

\textit{Research Question:} Does SemR-p demonstrate statistically significant sensitivity to performance differences between keyphrase models and across datasets?

\textit{Method:} We performed an Analysis of Variance (ANOVA) on the per-document SemR-p scores (using $k=3$), assessing the effects of \texttt{model} (Table~\ref{tab:models}), \texttt{dataset} (\textit{kp20k} vs. \textit{kptimes}), and their interaction. We calculated Eta-squared ($\eta^2$) effect sizes to quantify the proportion of variance attributable to each factor.
\paragraph{System-Level Ranking Agreement} To assess whether SemR-p produces interpretable rankings of system performance, we compared its model-level rankings to those induced by existing metrics.

\textit{Research Question:} Do the system-level rankings produced by SemR-p align with those produced by established metrics?

\textit{Method:} For each dataset, we computed the average metric score per model and derive rankings. We then computed Spearman rank correlations ($\rho$) between SemR-p and each baseline metric to quantify agreement.
\paragraph{Exploratory Factor Analysis} To investigate what underlying dimensions SemR-p aligns with, we conducted an exploratory factor analysis across all evaluation metrics.

\textit{Research Question:} What are the primary latent factors underlying the correlations among the selected keyphrase evaluation metrics, and how does SemR-p relate to these factors compared to other established metrics?

\textit{Method:} We performed EFA on standardised per-document scores from all 11 metrics. We retain two factors based on Parallel Analysis and applied Varimax rotation to aid interpretability. Factor loadings indicate each metric's association with the latent dimensions.
\paragraph{Qualitative Analysis} To better understand the unique scoring behaviour of SemR-p, we examine specific examples where it diverges significantly from baseline metrics.

\textit{Research Question:} In what practical scenarios does SemR-p's behaviour diverge from baseline metrics, and what do these cases reveal about its unique properties?

\textit{Method:} We selected representative cases from the evaluation corpus and presented the predictions, references, and associated metric scores. We then analysed scoring logic to highlight SemR-p’s unique strengths and limitations.

\section{Results}
\label{sec:results}

\subsection{Ablation Study of Parameter k}
\label{subsec:abl}

Our ablation study of the parameter $k$ revealed how this hyperparameter influences SemR-p's scoring behaviour and alignment with other metrics. We report three main findings, which guided our recommendation for a default value of $k$.

\begin{figure}[htbp!]
    \centering
    \includegraphics[width=0.8\linewidth]{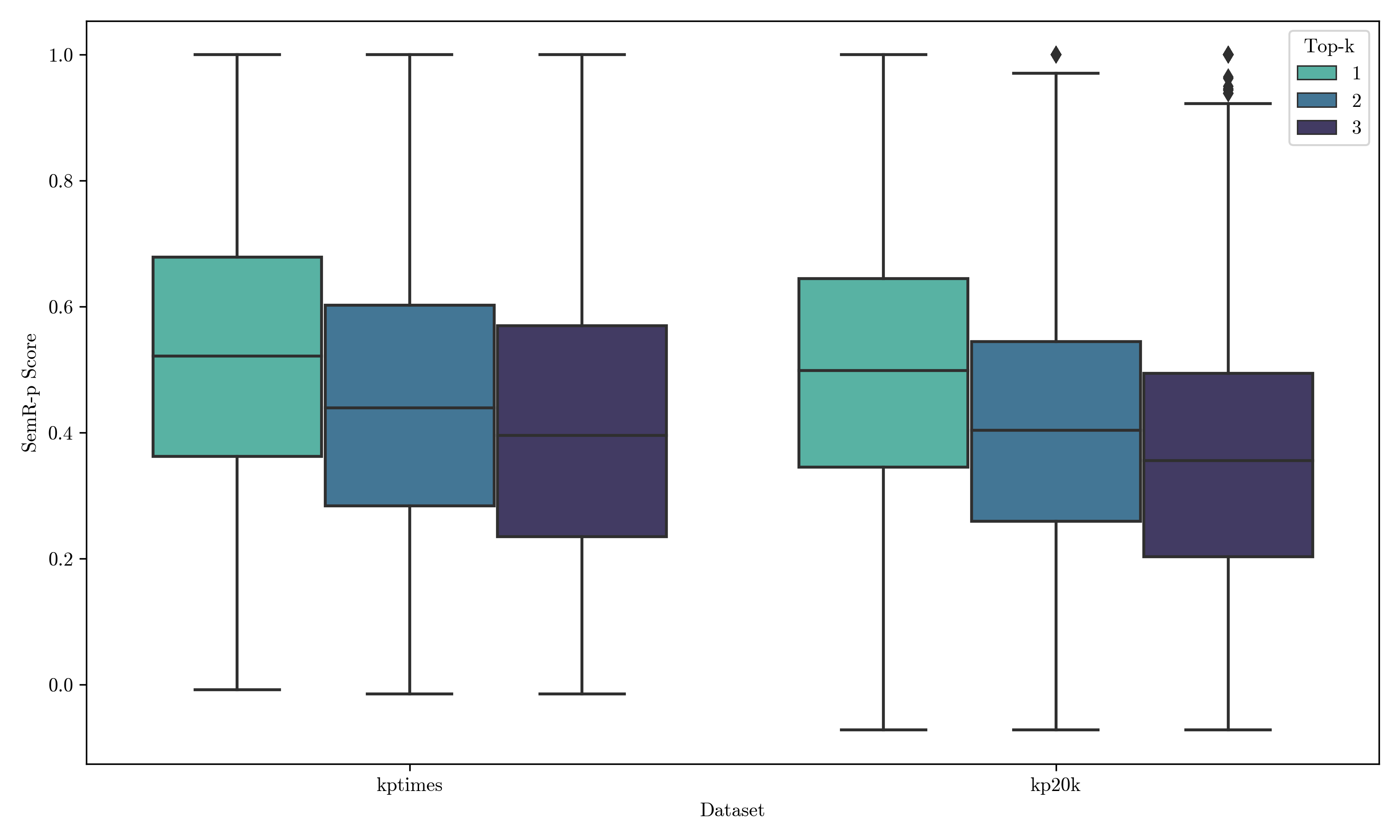}
   \caption{Distribution of \textit{SemR-p} scores on the two datasets as $k$ is varied.}
    \label{fig:ablation_k_distribution}
\end{figure}

\paragraph{Effect on Absolute Scores} We first examined how the choice of $k$ affects the raw scores produced by SemR-p, as visualised in Figure~\ref{fig:ablation_k_distribution}. As $k$ increased from 1 to 3, scores consistently decreased. For example, on the \textit{kp20k} dataset, the median score dropped from 0.50 to 0.35 ($p<0.001$). It appears that as more reference keyphrases are averaged in, they become progressively less semantically similar, lowering the average score.

The extent of this drop differs across domains. On \textit{kp20k} (scientific texts), the decrease was larger (medium effect size: Cliff's $\delta \approx 0.36$) than on \textit{kptimes} (news articles, $\delta \approx 0.29$). This reflects that scientific keyphrases are more semantically distinct, while news article keyphrases tend to be more redundant or overlapping \cite{jin-etal-2013-mining, gallina-etal-2019-kptimes}.
\FloatBarrier

\paragraph{Stability of Core Semantic Assessment} Despite changes in absolute values, SemR-p's core behaviour remained stable. Across all $k$ values, it maintained strong agreement with other semantic metrics like SemF1 and BERTScore (Spearman $\rho$ between 0.6 and 0.85). While some differences were statistically significant, the practical effect sizes were minimal (Cliff’s $\delta < 0.12$), suggesting these changes are not meaningful in practice. This important finding confirms that the core semantic assessment of SemR-p is robust and not fundamentally altered by the choice of $k$; the metric consistently agreed with other semantic measures on which systems perform better, regardless of how many references are used for averaging.

\begin{figure}[htbp!]
    \centering
    \includegraphics[width=0.6\linewidth]{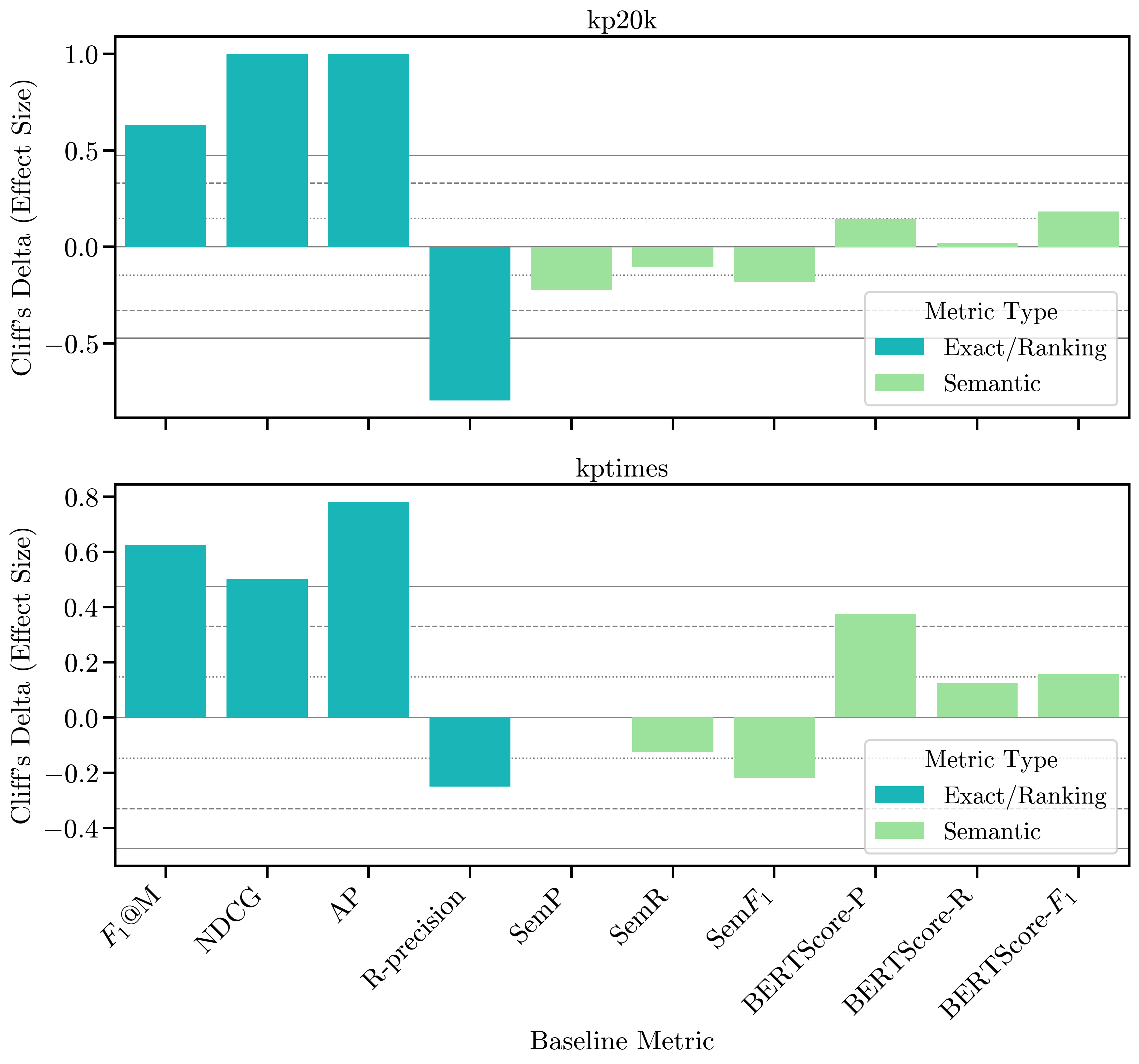}
    \caption{Change in SemR-p's Spearman correlation with baseline metrics when increasing the parameter $k$ from 1 to 3. The two plots show results for the \textit{kp20k} (top) and \textit{kptimes} (bottom) datasets. Each bar represents a baseline metric, and its height corresponds to the Cliff's Delta ($\delta$) effect size, indicating the magnitude and direction of the change. A positive $\delta$ value signifies that the correlation is stronger with $k=3$ compared to $k=1$, while a negative value signifies a weaker correlation.}
    \label{fig:k_sensitivity}
\end{figure}

\paragraph{Shifting Alignment with Non-Semantic Metrics} More interestingly, changing $k$ altered how SemR-p aligned with non-semantic metrics. While its correlation with semantic metrics was stable, altering $k$ from 1 to 3 induced large and statistically significant changes in its correlation profiles with exact-match and ranking metrics (Cliff's $\delta > 0.47$), as shown in Figure~\ref{fig:k_sensitivity}. Specifically:
\begin{itemize}
    \item At $k=3$, SemR-p showed stronger alignment with metrics such as F1@M, NDCG, and AP, which focus on overall set quality. One possible explanation is that averaging over more references ($k=3$) may make the semantic score reflect a broader portion of the reference keyphrase set, which could in turn increase its similarity to rank-based metrics. However, this remains a hypothesis rather than a confirmed causal relationship.
    \item At $k=1$, SemR-p relies on the single best semantic match. This does not necessarily yield high scores, but it can produce relatively higher values than when averaging across multiple references. Likewise, the approximate-matching strategy of R-Precision is lexically permissive, granting full credit for loose overlaps. This relative leniency in both metrics may help explain their divergence in correlation profile as $k$ increases from 1 to 3 for \textit{kp20k}, which presents more semantically heterogeneous and specialised reference sets.
\end{itemize}

This shift reflects a trade-off: $k=1$ emphasises best-match quality, while higher $k$ values reward broader semantic alignment with the reference set.

\paragraph{Recommendation for $k$} Taken together, these analyses show that the parameter $k$ acts as a tuner for the metric's semantic focus: 
\begin{itemize}
    \item $k=1$ offers a fine-grained, precision-oriented evaluation focused on the best possible match for each prediction;
\item $k=3$ seems to better capture the holistic retrieval quality of the top-$R$ set.
\end{itemize}

Because $k=3$ produced stronger alignment with a wider range of metrics, lexical, ranking-based and semantic, and offers a more balanced evaluation, we recommend using $k=3$ as the default setting. 

\FloatBarrier

\subsection{Sensitivity to Model Differences}
\label{sec:anova}

We next tested whether SemR-p could reliably distinguish between different models and domains. A two-way ANOVA on over 48,000 document-level scores confirmed that model identity had a significant and substantial effect on SemR-p scores ($F[7, 47984] = 1353.91$, $p < 0.001$). Specifically, the choice of model explained 16\% of the total variance in scores ($\eta^2 = 0.16$), which is considered a large effect in this context. By comparison, the dataset effect ($\eta^2 = 0.008$) and the model–dataset interaction ($\eta^2 = 0.021$) were also statistically significant, but their practical impact was small. This is expected, as document-level variability accounted for most of the variance (\(\approx 81\%\)). These results confirm that SemR-p captures 
differences between keyphrase systems, validating its sensitivity to model performance across domains.
\subsection{System-Level Ranking Agreement}
\label{sec:rankagr}

\begin{figure}[htbp!]
    \centering
    \includegraphics[width=0.8\linewidth]{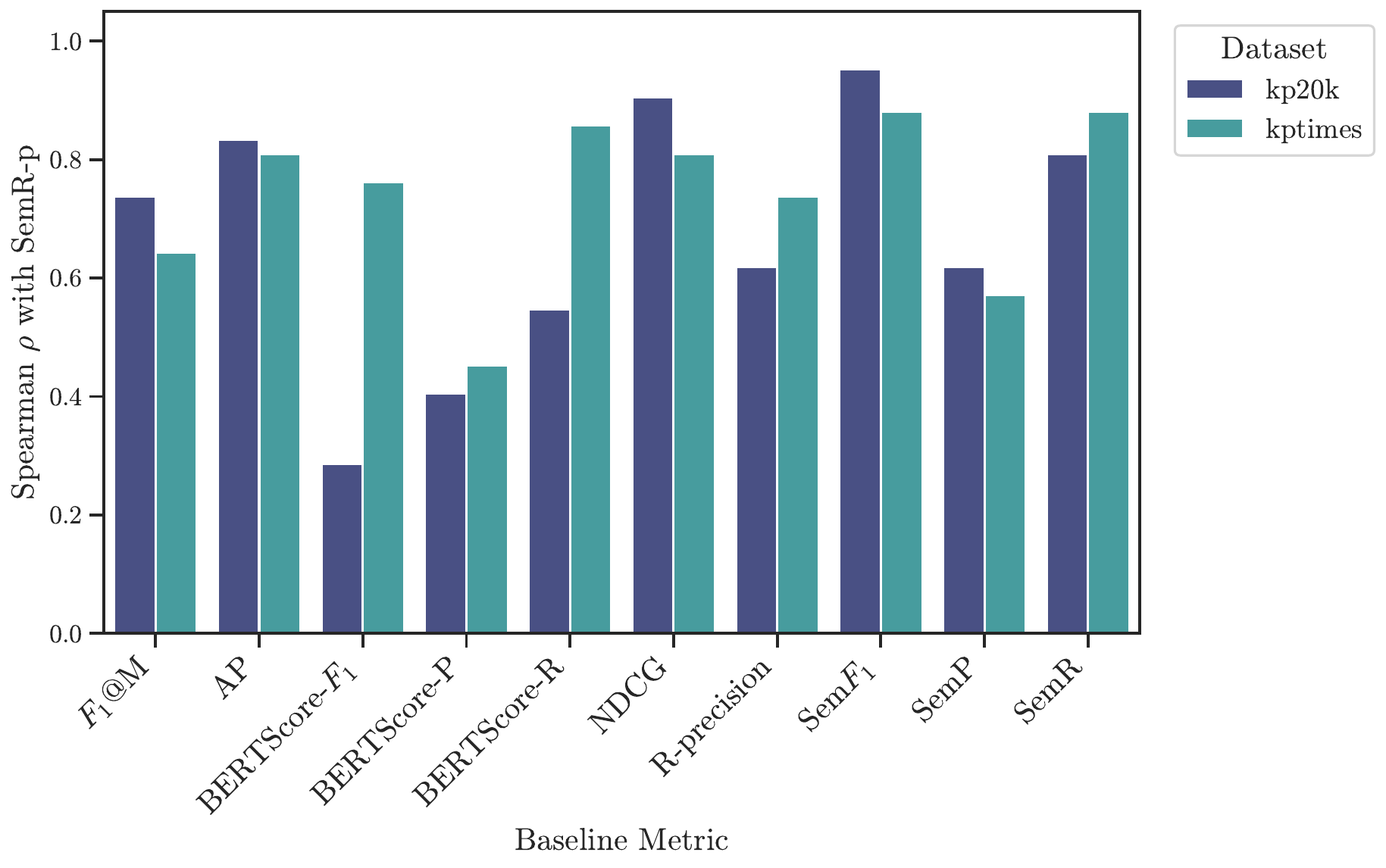}
    \caption{Spearman rank correlation ($\rho$) between system rankings produced by SemR-p ($k=3$) and those produced by each baseline metric, shown separately for the \textit{kp20k} and \textit{kptimes} datasets. Each bar reflects how similarly SemR-p ranks systems compared to the respective baseline. Higher values indicate stronger agreement.}
    \label{fig:sys-agr}
\end{figure}

We now examine how closely SemR-p's overall rankings of keyphrase systems agreed with those produced by existing evaluation metrics. To do this, we compared the average scores assigned to each system by SemR-p and by each baseline metric, and calculated how strongly their rankings match using Spearman correlation ($\rho$). This analysis gave us a broad sense of whether SemR-p tends to judge system performance in a similar way to well-established methods (Figure~\ref{fig:sys-agr}).

Overall, SemR-p showed positive correlation with all baselines, indicating that it captures meaningful and widely accepted aspects of system quality. However, it showed the weakest alignment with the BERTScore variants. This divergence might stem from several methodological differences: BERTScore compares tokens rather than phrases, uses a different embedding model, lacks rank sensitivity, and aggregates scores differently. These differences might make it less directly comparable to SemR-p, which operates at the phrase level and incorporates a ranking structure.

Dataset characteristics also play a role: correlations were generally higher on the \textit{kptimes} dataset than on \textit{kp20k}, particularly for BERTScore. This suggests that SemR-p's hybrid evaluation style becomes more distinctive in technical domains like scientific writing, where vocabulary and phrasing are highly specialised.

Taken together, these results characterise SemR-p as a balanced and nuanced metric that aligns well with established semantic and ranking metrics.

\FloatBarrier

\subsection{Exploratory Factor Analysis}
\label{sec:FA}

\begin{figure}[htbp!]
    \centering
    \includegraphics[width=0.7\linewidth]{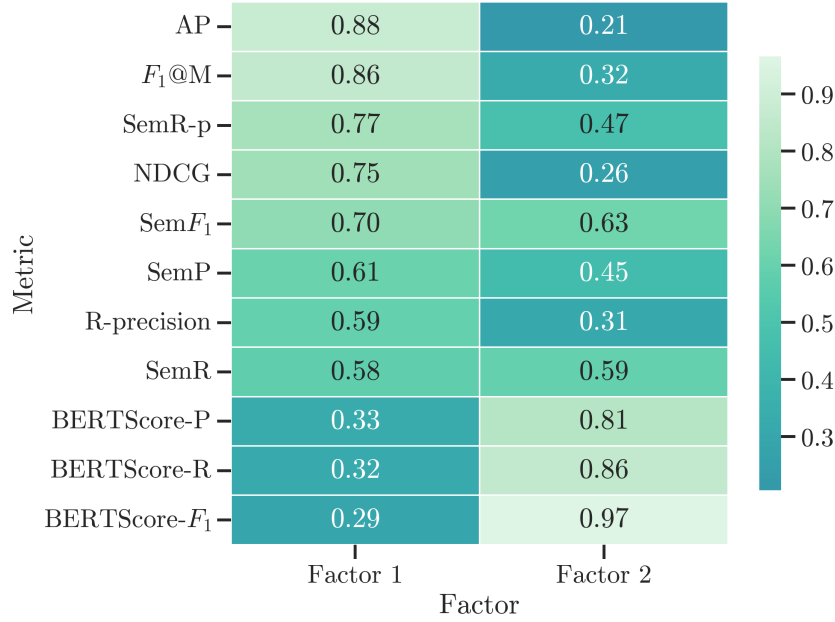}
    \caption{Heatmap of factor loadings from Exploratory Factor Analysis with two factors. Loadings indicate the correlation between each evaluation metric and the two extracted latent factors.}
    \label{fig:FA}
\end{figure}

To uncover broader structural patterns in the metric landscape beyond pairwise correlations, we conducted an Exploratory Factor Analysis to identify latent dimensions underlying their relationships. Using Parallel Analysis to determine the appropriate number of factors, we retained two latent dimensions, which together explain 75.4\% of the shared variance across the 11 evaluated metrics (Factor 1: 41.0\%, Factor 2: 34.4\%).  This two-factor solution offers a clear and interpretable structure for understanding the metric landscape. The contribution of each metric to each factor, factor loading, is shown in Figure~\ref{fig:FA}.

Factor 1 can be characterised as 'General Effectiveness \& Ranking'. It includes high loadings from classic evaluation metrics focused on exact matches (F1@M), ranked retrieval quality (AP, NDCG), and precision (\textit{R-Precision}). Several semantic metrics, including SemR-p, also load heavily onto this factor, suggesting that it reflects a broad notion of system performance across both traditional and semantic criteria.

Factor 2 represents 'Contextual Semantic Similarity', being almost exclusively dominated by the three BERTScore variants. This suggests a distinct dimension related to the specific type of deep contextual semantic matching operationalised by BERTScore.

The analysis suggests that SemR-p may function as a bridging metric, showing substantial loadings on both Factor 1 (0.77) and Factor 2 (0.47) and therefore relating to both traditional effectiveness and semantic similarity. While other semantic metrics (e.g., SemF1, SemR) also display this dual pattern, SemR-p has the strongest association with Factor 1 among them, consistent with its R-Precision-based design, which emphasises ranked retrieval quality. Its moderate loading on Factor 2 indicates that it incorporates semantic information while retaining its rank-aware behaviour.

\FloatBarrier
\subsection{Qualitative Analysis}
\label{sec:QA}

To provide practical intuition beyond the aggregate statistics, we qualitatively analysed selected examples. The goal was to present diagnostic contrast cases that illustrate SemR-p's specific properties and behaviours by comparing it against key baseline metrics. The full set of contrast cases, metric scores and reference and predictions sets are available in Table~\ref{tab:qualitative_cases}.

\paragraph{Semantic Reward vs. Exact Match Failure}
In Case 1, contrasting SemR-p and F1@M, there was a single reference keyphrase. While the model predicted ten, none were an exact match. Consequently, F1@M scored 0.0. SemR-p, however, assigned a high score of 0.71 by recognising the strong semantic similarity between the prediction and the target. This suggests that SemR-p can partially account for relevant lexical variations where strict matching fails.

\paragraph{Rank Sensitivity vs. Recall Focus}
Case 2 contrasted SemR-p and F1@M again. The model predicted all references correctly, but only at ranks 4, 5, and 6. Because all references were found within the top 10, F1@M was moderate (0.46). In contrast, SemR-p was very low (0.06) as it evaluated only the poor top $R=3$ predictions. This indicates that SemR-p tends to penalise models that do not place relevant keyphrases near the top.

\paragraph{Top-Rank Focus vs. Global Semantic Assessment}
Case 3 contrasted SemR-p with BERTScore-F1. SemR-p assigned a perfect score of 1.0, as all $R=3$ references were matched in the top 3 ranks. BERTScore-F1, however, was considerably lower (0.57). This is because BERTScore evaluates all ten predictions globally, and the lower-ranked predictions were deemed semantically dissimilar to the core targets, reducing the overall score. This case highlights the different focus of SemR-p on top-ranked relevance versus BERTScore's more holistic, rank-agnostic assessment.

\paragraph{Semantic Assessment Divergence}
A different comparison with BERTScore-F1 highlighted how the specific semantic assessment method can lead to divergent scores. In Case 4, the model produced five complex but highly relevant predictions against a large set of 18 references. BERTScore-F1 assigned a high score of 0.75, effectively capturing the strong overall semantic overlap. SemR-p, however, was very low (0.19). Despite one exact match, its phrase-level semantic component assigned very low scores to the other four composite phrases, suggesting a stricter or different handling of phrase compositionality compared to BERTScore's token-level approach.

\paragraph{Semantic Near-Miss vs. Approximate Matching}
Case 5, which compared SemR-p and R-Precision, had two references: \texttt{fuzzy topology} and \texttt{fuzzy strongly semiopen set}. The top prediction was \texttt{fuzzy less strongly semiopen set}. The baseline R-Precision scored 0.0 because the word \texttt{less} prevented a match under its approximate (e.g., morphological, substring) matching rules. SemR-p, however, assigned a high score of 0.66. This suggests that SemR-p can reward conceptually close predictions that lexical matchers may miss.

\paragraph{Specificity vs. Approximate Matching}
Case 6 shows another comparison with R-Precision, in which the references included \texttt{mobile ad hoc networks} and \texttt{quality information delivery}. The model's top three predictions were the general, single-word terms \texttt{mobile}, \texttt{information}, and \texttt{networks}. The baseline R-Precision, using an approximate (substring) matching rule, assigned a perfect score of 1.0 because each prediction is a substring of a reference. SemR-p, however, assigned a low score of 0.25. Its semantic component assigned low similarity scores, penalising the lack of specificity.

\paragraph{Rank-Aware Semantic Reward}
Case 7 compares SemR-p with NDCG. The model produced several semantically strong but non-lexical matches in the top ranks. Because no top predictions were exact stem matches, NDCG (which relies on exact stem matches for relevance in our baseline) scored 0.0. SemR-p, however, assigned a high score of 0.636, rewarding the strong semantic similarity of these well-ranked predictions, showing that it can incorporate semantic relevance in a rank-sensitive manner.

\paragraph{Stricter Semantic Assessment}
Finally, Case 8 shows another comparison with NDCG. The model's top prediction was an exact match to a reference, leading to a perfect NDCG score of 1.0. SemR-p also gave this prediction a full score of 1.0. However, the subsequent five predictions (e.g., \texttt{event based modeling}, \texttt{multimedia databases}), while topically related, were assigned very low or even negative semantic scores on average by SemR-p's semantic component. This pulled the final SemR-p score down to 0.12, illustrating a case where its specific phrase-level semantic assessment can be stricter than a simple exact-match-based ranking metric if follow-up predictions are not deemed sufficiently close. This effect may have been amplified by the large and diverse reference set ($R=34$).

\begin{table*}[htbp!]
\centering
\caption{Data for Qualitative Case Studies Illustrating Divergences between SemR-p and Baseline Metrics.}
\label{tab:qualitative_cases}
\resizebox{\textwidth}{!}{%
\begin{tabular}{p{0.4cm} p{4.0cm} p{5.5cm} p{7.5cm}} 
\toprule
\textbf{Case} & \textbf{Score Comparison} & \textbf{Reference(s)} & \textbf{Prediction(s)} \\
\midrule
1 & SemR-p (0.71) > F1@M (0.0) & (1) mobility model & (10) mobility; pedestrian mobility; observations; train station; pedestrian; train; station; modeling pedestrian; modeling pedestrian mobility; train station scenarios\\
\addlinespace
2 & SemR-p (0.06) < F1@M (0.46) & (3) temporal logic; multi agent systems; model checking & (10) expressiveness; complexity; atl; temporal logic; multi agent systems; model checking; alternating transition systems; concurrent games; bisimulation; modal logic \\
\addlinespace
3 & SemR-p (1.0) > BERTScore-F1 (0.57) & (3) grounds for sculpture; art; sculpture & (10) art; sculptural; grounds for sculpture; jon isherwood; foon sham; john ruppert; wendy ross; enclosures; the grounds for sculpture; mr. sham \\
\addlinespace
4 & SemR-p (0.19) < BERTScore-F1 (0.75) & (18) approximation; size; weakly connected dominating set; connection; dominating set; clustering; ... & (5) minimum size weakly connected dominating sets; clustering mobile ad hoc networks; approximation algorithms; connected dominating sets; graph connectivity degradation \\
\addlinespace
5 & SemR-p (0.66) > R-Precision (0.0) & (2) fuzzy topology; fuzzy strongly semiopen set & (4) fuzzy less strongly semiopen set; fuzzy less strong semicontinuity; fuzzy topology; fuzzy semiopen set \\
\addlinespace
6 & SemR-p (0.25) < R-Precision (1.0) & (3) mobile ad hoc networks; optimal stopping theory; quality information delivery & (10) mobile; information; networks; delivery; mobile context aware applications; quality contextual information; policies; data consumption; quality aware scheduling; reporting \\
\addlinespace
7 & SemR-p (0.636) > NDCG (0.0) & (4) blind watermarking; multiple watermarking; wavelet packets; parameterized wavelet filters & (6) wavelet based blind watermarking; wavelet domain transformations; wavelet filter parametrization; wavelet packet decomposition; secret watermark embedding; image distribution chain\\
\addlinespace
8 & SemR-p (0.120) < NDCG (1.0) & (34) event; model; process; digital media; informal; video; image; ... & (6) digital media; event based modeling; event based processing; digital media processing; multimedia databases; multimedia processing \\
\bottomrule
\end{tabular}%
}
\end{table*}
\section{Discussion}
\label{sec:discussion}

The central challenge in keyphrase evaluation is to design metrics that better reflect what humans perceive as meaningful, relevant, or necessary to complete their task \cite{Cai_Leckner_Björklund_2025}. Semantic R-Precision (SemR-p) was introduced as a step towards bridging this gap. Our analyses aimed to go beyond basic validation, highlighting how specific design choices may support a more nuanced, human-aligned evaluation.

\paragraph{The Interplay of Ranking and Semantics}
Our results indicate that considering ranking and semantics as fully separable evaluation dimensions may be insufficient. Factor analysis suggests that 'General Effectiveness \& Ranking' and 'Contextual Semantic Similarity' emerge as two dominant, yet distinct, latent factors in keyphrase evaluation. SemR-p appears to align more closely with the 'Ranking' factor compared to other semantic metrics like SemF1, suggesting that it maintains a degree of rank sensitivity while incorporating semantic flexibility. This tendency is illustrated in the qualitative analysis: SemR-p, similarly to NDCG, penalises poorly ranked predictions (Case 2), but unlike NDCG, it can also assign higher scores to semantically strong, well-ranked near-misses (Case 7). These examples suggest that SemR-p may capture the interaction between relevance and ranking in a manner that aligns with a user-centric view of keyphrase quality.

\paragraph{The Role of the Semantic Neighbourhood ($k$)}
The ablation study (Section~\ref{subsec:abl}) indicates that the parameter $k$ influences the metric’s emphasis on different types of semantic relevance. $k=1$ tends to reward a prediction if it strongly matches at least one core concept, consistent with a precision-oriented judgment, while larger values (e.g., $k=3$) appear to favour a broader assessment of how predictions fit into the semantic space of the reference set. The domain-dependent impact of $k$ suggests that the semantic structure of the reference set can interact with the metric’s behaviour. While $k=3$ may serve as a reasonable default due to its balanced alignment with several baseline metrics, further exploration across diverse domains could provide guidance on tailoring $k$ for more context-specific evaluation.

\paragraph{Implications for Keyphrase Model Evaluation}
These findings suggest that relying on a single metric may provide an incomplete picture of model performance. A model that performs well on recall-oriented metrics like F1@M or SemF1 may still rank relevant keyphrases suboptimally, a limitation that SemR-p is designed to highlight. Conversely, models that excel on lexically-bound ranking metrics like NDCG might be penalised for generating semantically appropriate paraphrases. By integrating semantic similarity with rank sensitivity, SemR-p offers a more balanced perspective for benchmarking, encouraging the development of models that consider both semantic relevance and proper ranking of predictions.

\section{Limitations and Future Work}
\label{sec:limitations}

While SemR-p is supported by both empirical results and conceptual reasoning, several limitations and open questions remain, suggesting avenues for future investigation.

First, SemR-p has not yet been validated against direct human judgments of keyphrase quality. While it aligns with well-established cognitive principles, such as semantic relevance and rank sensitivity, future work will assess its correlation with human evaluations to confirm its effectiveness in practical annotation, behavioural study or user-facing scenarios.

Second, the semantic scoring component of SemR-p depends on the chosen embedding model. In this study, we adopted a domain-tuned Sentence Transformer, but evaluating alternative or more recent embedding architectures could yield insights into robustness and further clarify performance differences with metrics like BERTScore. Similarly, the implementation of semantic scoring could be refined: while mean pooling and cosine similarity offer a strong baseline, alternative pooling mechanisms such as max, attention-based, or weighted pooling may offer more nuanced representations, though often at higher computational cost \cite{XingetalPooling}.

Third, the metric’s similarity aggregation method relies on the parameter $k$, which determines how many reference keyphrases are considered when computing semantic similarity. Although our ablation study found $k=3$ to be a robust default, future work may explore strategies to select $k$ dynamically. For instance, $k$ could scale with the number of references ($R$), or adapt to the semantic density of the reference set estimated through the distribution of pairwise cosine similarities, via their average or entropy. However, such methods require calibration and may introduce computational complexity. Structural analyses such as factor analysis at different $k$ values ($k=1$) could further illuminate how the metric’s behaviour shifts between precision-focused and recall-oriented profiles.

Finally, SemR-p does not explicitly penalise semantic redundancy: multiple predicted phrases can receive partial credit for aligning with the same reference. While this behaviour reflects the ambiguity in keyphrase annotation, it limits the metric’s ability to assess conceptual diversity. A promising extension would be a diversity-aware variant (\textit{SemR-p-Div}), which incorporates one-to-one matching, for example via maximum bipartite matching. This approach, based on the assignment problem \cite{munkres1957algorithms}, has been used in semantic set matching tasks \cite{Mundra2023KOIOS}, and would reward predictions that align with distinct reference concepts. It could also serve as an adaptive mechanism for selecting $k$. Such a variant would provide a useful complement to existing reference-free diversity metrics \cite{wu-etal-2024-kpeval} and extend SemR-p’s ability to assess not just relevance, but conceptual breadth.

Despite these considerations, SemR-p offers a significant contribution by providing a more holistic and principled assessment than metrics focusing purely on lexical matches, ranking without semantics, or rank-agnostic semantic similarity. It enables evaluation based on a system's ability to place semantically relevant keyphrases at the top of its predictions, aligning closely with practical application needs and offering a clearer lens through which to view progress in keyphrase prediction.
\section{Conclusion}
\label{sec:conclusion}

In this work, we introduced Semantic R-Precision (SemR-p), a novel metric for keyphrase evaluation designed to better align with human judgments by integrating semantic relevance with rank awareness. Built upon the R-precision framework to model user focus on top results, SemR-p rewards unambiguous exact stem matches while using embedding-based semantic similarity to credit conceptually relevant but lexically different predictions.

Through a comprehensive suite of experiments, including ablation studies, statistical validation, system-level ranking comparisons, factor analysis, and qualitative case studies, we demonstrated SemR-p's properties. The results suggest that SemR-p is sensitive to differences in model performance and can help bridge traditional evaluation dimensions (ranking and exactness) with modern semantic similarity measures. Its factor loading profile, particularly its tendency to align with general effectiveness while retaining semantic sensitivity, highlights distinctions from existing metrics.

Overall, SemR-p provides researchers and practitioners with a more nuanced and interpretable tool for evaluating keyphrase systems. By accounting for both semantic relevance and ranking quality, it can complement existing metrics and offer additional insight into model performance in scenarios where both dimensions are important. 
\begin{acknowledgments}
This work was supported by the Ministry of Science, Research and the Arts of Baden-Württemberg (MWK) and KIT's Accessibility through AI-based Assistive Technology (KATE) Graduate School.
\end{acknowledgments}

\section*{Declaration on Generative AI}
During the preparation of this work, the author(s) used Gemini in order to rephrase, proofread or summarise text content. After using this tool/service, the author(s) reviewed and edited the content as needed and take(s) full responsibility for the publication’s content.

\FloatBarrier

\bibliography{mybibfile}

\end{document}